\documentclass[preprint,nofootinbib,aps,superscriptaddress]{revtex4}

\newcommand{\beq}{\begin{equation}}
\newcommand{\eeq}{\end{equation}}
\newcommand{\beqa}{\begin{eqnarray}}
\newcommand{\eeqa}{\end{eqnarray}}
\newcommand{\no}{\nonumber}

\begin{document}

\preprint{\vbox{\hbox{LBNL--53794} \hbox{WIS/25/03-Oct-DPP}
  \hbox{SLAC-PUB-10199} \hbox{hep-ph/0310242}}}

\vspace*{1.25cm}

\title{\boldmath Comment on extracting $\alpha$ from $B\to\rho\rho$}

\author{Adam F.\ Falk}\email{falk@jhu.edu}
\affiliation{Department of Physics and Astronomy, The Johns Hopkins
  University\\ 3400 North Charles Street, Baltimore, MD 21218, USA}

\author{Zoltan Ligeti}\email{zligeti@lbl.gov}
\affiliation{Ernest Orlando Lawrence Berkeley National Laboratory \\
  University of California, Berkeley, CA 94720, USA}

\author{Yosef Nir}\email{yosef.nir@weizmann.ac.il}
\affiliation{Department of Particle Physics \\
  Weizmann Institute of Science, Rehovot 76100, Israel}

\author{Helen Quinn}\email{quinn@slac.stanford.edu}
\affiliation{Stanford Linear Accelerator Center\\
  Stanford University, Menlo Park, CA 94025, USA\\[-12pt] $\phantom{}$}

%\date{\today}
%\pacs{13.25.Hw, 11.30.Er, 12.15.Hh}

\begin{abstract}

Recent experimental results on $B\to\rho\rho$
decays~\cite{Zhang:2003up,Aubert:2003mm,Aubert:2003zv} indicate that the CP
asymmetry $S_{\rho^+\rho^-}$ will give an interesting determination of
$\alpha=\arg\left[-(V_{td}V_{tb}^*)/(V_{ud}V_{ub}^*)\right]$.  In the limit
when the $\rho$ width is neglected, the $B\to\pi\pi$ isospin analysis can also
be applied to $B\to\rho\rho$, once an angular analysis is used to separate
transversity modes.  The present bound on the shift of $S_{\rho^+\rho^-}$ from
the true $\sin2\alpha$ is already stronger than it is for $S_{\pi^+\pi^-}$.  We
point out a subtle violation of the isospin relations when the two $\rho$
mesons are observed with different invariant masses, and how to constrain this
effect experimentally.

\end{abstract}

\maketitle

%%%%%%%%%%%%%%%%%%%%%%%%%%%%%%%%%%%%%%%%%%%%%%%%%%%%%%%%%%%%%%%%%%%%%%
\section{Introduction}
\label{sec:introduction}

Rates and polarization fractions for various $B\to\rho\rho$ decays have been
recently measured~\cite{Zhang:2003up,Aubert:2003mm,Aubert:2003zv}. First
measurements of CP asymmetries in these modes are expected in the near future.
This note is a brief comment on the application of isospin analysis to these
modes, similar to that for $\pi\pi$ channels~\cite{Gronau:1990ka} to extract
Standard Model parameters, and in particular the CKM phase  $\alpha \equiv
\phi_2 \equiv \arg\left[-\left(V_{td}V_{tb}^*\right) /
\left(V_{ud}V_{ub}^*\right)\right]$, from these measurements. It is important
to constrain such CKM phases as precisely as possible in many independent ways.
Inconsistent results from different approaches could be an indicator of new
physics as various measurements that are related in the Standard Model can be
affected differently by possible contributions from physics beyond the Standard
Model. Here we comment on the need to parameterize the data to allow for the
impact of possible $I=1$ contributions that can occur if the two $\rho$ mesons
have different masses.\footnote{By $\rho$ mass we mean throughout this paper
the invariant mass of the pion pair from the decay of that $\rho$.}

In the standard parameterization for the CKM matrix, the phase dependence of
the $B\to\rho^i\rho^j$ decay amplitudes can be written as
\beqa\label{defpt}
A_{ij}&=&T_{ij}e^{+i\gamma}+P_{ij}e^{-i\beta},\no\\
\bar A_{ij}&=&T_{ij}e^{-i\gamma}+P_{ij}e^{+i\beta},
\eeqa
where $A_{ij}$ describe $B^+$ and $B^0$ decays, $\bar A_{ij}$ describes $B^-$
and $\overline{B}{}^0$ decays, and $\beta$, $\gamma$ (and
$\alpha=\pi-\beta-\gamma$) are the angles of the unitarity triangle (for their
precise definitions, see {\it e.g.}, Ref.~\cite{Babarbook}).  $T_{ij}$ is
dominated by the tree diagram, while $P_{ij}$ comes primarily from so-called
penguin diagrams. An important role in the CP asymmetries in neutral $B$ decays
is played by the $B^0-\overline{B}{}^0$ mixing amplitude, which has the
following CKM phase dependence:
\beq\label{mixamp}
M_{12}=|M_{12}|e^{2i\beta}.
\eeq
The dominant CP violating effect in the $B\to\rho^+\rho^-$ decay comes from the
interference between the $B^0-\overline{B}{}^0$ mixing amplitude and $T_{+-}$.
As can be deduced from Eqs.~(\ref{defpt}) and (\ref{mixamp}), this effect is
sensitive to the phase $\alpha = \pi -\beta -\gamma$ (or $\phi_2 =\pi -\phi_1
-\phi_3$).

The time dependent CP asymmetry in $B\to\rho^+\rho^-$ can be parametrized as
follows:
\beq\label{defscpm}
\frac
{\Gamma(\overline{B}{}^0_{\rm phys}(t)\to\rho^+\rho^-)-\Gamma({B^0_{\rm
      phys}(t)}\to\rho^+\rho^-)}
{\Gamma(\overline{B}{}^0_{\rm phys}(t)\to\rho^+\rho^-)+\Gamma({B^0_{\rm
      phys}(t)}\to\rho^+\rho^-)}
= S_{+-}\sin(\Delta m\,t)-C_{+-}\cos(\Delta m\,t)\,.
\eeq
If $|P_{+-}/T_{+-}|$ were zero, so that a single weak phase dominates the decay
and if, in addition, the final state were purely CP even, then
$S_{+-}=\sin2\alpha$ (and $C_{+-}=0$).  A separation of final CP eigenstates is
possible with angular analysis~\cite{Dunietz:1990cj}; as we will see below the
data show that the decays to charged $\rho$'s are dominantly longitudinally
polarized and thus CP even.

Other CP violating effects in $B\to\rho\rho$ decays arise from the interference
between the $T$ and $P$ terms in Eq.~(\ref{defpt}) or from interference between
mixing and $P$ amplitudes. These effects not only have different weak phase
dependences, but also depend on the amplitude ratio $|P/T|$ and the strong
phase $\arg(P/T)$. These complicate the relationship between the measured CP
violation and the phase $\alpha$. For a given final transversity $\sigma$ (see
discussion below), this more complicated relation can be parametrized as
follows:
\beq\label{comrel}
S^\sigma_{+-} = \sqrt{1-(C^\sigma_{+-})^2}\, \sin(2\alpha+2\delta_\sigma) \,.
\eeq
In the case of two pions, Gronau and London~\cite{Gronau:1990ka} showed how to
use the six flavor-tagged $B\to\pi\pi$ rates and isospin symmetry to precisely
determine $\alpha$ even in the presence of the additional CP violating effects.
Later work showed how one can use the isospin relations to bound the
uncertainties in $\alpha$, even when sufficient data to complete the full
analysis is not available~\cite{Grossman:1997jr, Charles:1998qx,
Gronau:2001ff}. These methods can be applied also for the decays to two $\rho$
mesons. The current experimental data implies that the $B\to\rho\rho$ case will
give a better intermediate result.

%%%%%%%%%%%%%%%%%%%%%%%%%%%%%%%%%%%%%%%
\section{Bose Statistics and Broad Resonances}

The vector-vector decays of a spin zero $B$ meson can have orbital angular
momentum $L=0$, 1, or 2. Hence, for two vector particles, they include both
even and odd CP modes. Since the decaying $B$ meson is spin 0, the total spin
of the two vector mesons must be equal to and oppositely aligned to the orbital
angular momentum, $L$. Thus, in the case of two identical vector mesons, such
as two equal mass $\rho$ mesons, independent of the value of $L$, the combined
space plus spin wave function of the the two identical vector mesons is
symmetric under particle exchange. Bose statistics then tells us that, just as
in the case of two pions, the isospin of the two $\rho$ meson state must be
symmetric under exchange of the particles, thereby eliminating any possible
$I=1$ contributions.\footnote{In Section 6.1.2.2 of the Babar Physics
Book~\cite{Babarbook} this argument is given correctly for the $L=0,2$ case.
However, an incorrect conclusion that isospin analysis is not possible for the
$L=1$ component is stated.  Mea culpa, HQ.}

While the above argument is made in terms of the amplitudes of a given $L$, it
applies for all $L$. Thus it is equally valid when applied to the amplitudes
expressed in any alternative angular decomposition. The set of basis functions
for describing the decays used in the experimental analyses are labelled by the
transversity $\sigma = 0, \parallel, \perp$ of the $\rho$ mesons (which both
must be the same since the initial state has spin zero). Thus, from this point
on, our discussion will be in this basis. Note that once this basis is chosen
there is no longer any sense in which one can separate the different orbital
angular momentum contributions within a given transversity-labelled state.
Since transversity-labelled amplitudes are a choice of three orthogonal angular
basis functions for analyzing the decays, they contain the full angular
momentum information. Thus we have a complete set of amplitudes,
$A^{\sigma}_{ij}=A[B\to(\rho^i\rho^j)_\sigma]$, where $\sigma$ is the
transversity label and $i$ and $j$ are the charges of the two $\rho$ mesons.
The CP of a given transversity state is well-defined, in the case at hand the
states $\sigma=0$ and $\parallel$ are CP even, while the $\sigma =\perp$ states
are CP odd~\cite{Dunietz:1990cj}.

The above arguments for the absence of $I=1$ in each transversity state do not
apply for general four-pion amplitudes. This contribution exists even when two
pion pairs have the same invariant mass and angular momentum. Indeed the fact
that $\rho$ mesons have a significant width reintroduces the possibility of
$I=1$ contributions even for a pair of longitudinally polarized $\rho$
particles. In each $B\to \rho\rho$ event the invariant mass of each $\rho$ is
measured, and the two values can differ by an amount of order of $\Gamma_\rho$,
or rather by the width of the region allowed by experimental cuts on the data. 
The $B\to \rho\rho$ amplitude for two $\rho$ mesons with charges $q_1,\ q_2$,
masses $m_1,\ m_2$ and helicities $\lambda_1=\lambda_2$ can have a part which
is antisymmetric under the interchange of the values of $m_1$ and $m_2$, and
thus, by Bose statistics, this amplitude is also antisymmetric in the combined
(space, spin, isospin) wave function, thus allowing odd isospin, despite the
fact that $L=S$. In contrast, the dependence of the even-isospin amplitudes on
the $\rho$ masses is symmetric under interchange of $m_1$ and~$m_2$. The
different isospin amplitudes do not interfere.  Our main point in this note is
that the fits to data should explicitly include the possibility of the
odd-isospin contribution in $B\to\rho\rho$.

The size of the $I=1$ contribution is a dynamical question; we make no
prediction.  We cannot rule out the presence of $I=1$ contributions of order
$(\Gamma_\rho/m_\rho)^2$ in the total rate.  The fact that this amplitude must
vanish for equal $\rho$ meson masses gives it a distinct distribution as a
function of $m_1$ and $m_2$ from the leading even-isospin terms. The leading
contribution to the rate due to the amplitude antisymmetric in $m_1$ and $m_2$
can be parameterized by adding to the fits a term of the form
\beq\label{fitterm}
\left[c\ \frac{m_1-m_2}{m_\rho}\right]^2
  \left|B_\rho(m_1^2)B_\rho(m_2^2)\right|^2 ,
\eeq
where $B_\rho(s)$ is the Breit-Wigner.  This
contribution vanishes where the even-isospin contribution peaks.
The $I=1$ contributions in the $\rho^+\rho^-$ and $\rho^\pm\rho^0$
channels are unrelated, while there is no such contribution to
$\rho^0\rho^0$.  Note that even-isospin contributions of the same
form are also possible, {\it e.g.}, from the cross-term in
\beq\label{evenI}
\left[a + b\ \frac{(m_1-m_2)^2}{m_\rho^2}\right]^2
  \left|B_\rho(m_1^2)B_\rho(m_2^2)\right|^2 .
\eeq
We expect $a,b$ and $c$ to be of the same order, so the even-isospin
contribution proportional to $ab$ could be comparable to the $I=1$ component.

The question is whether the extraction of the leading even-isospin
amplitudes [the $a^2$ term in Eq.~(\ref{evenI})] is sensitive
to possible contributions of the form (\ref{fitterm}). Independent
of whether the correction term is dominated by the $c^2$ term of
Eq.~(\ref{fitterm}) or the interference of $a$ and $b$ in
Eq.~(\ref{evenI}), the stability of the 
fit for the $a^2$ term can be tested. If the addition of terms of the
form (\ref{fitterm}) causes the value of the leading term to shift
significantly then further tests must be made to ensure a stable value
for the on-peak amplitudes. If adding such a term does not
significantly change the result for the leading term, then we can be
confident that the correct on-peak amplitudes have been measured.

While the $I=1$ contribution must be positive, the subleading even-isospin
contributions may have either sign. Thus, even if a fit to the data finds that
contributions to the rate of the form in Eq.~(\ref{fitterm}) are small, that
could still be due to cancellations. Such a cancellation would be accidental in
either the $\rho^+\rho^-$ or the $\rho^\pm\rho^0$ channels, and it is unlikely
to occur in both. Thus, if the fits in both of these modes are insensitive to
terms of the form (\ref{fitterm}), then it is probably safe to assume that the
$I=1$ contributions are likewise small. But, as we stress above, it is not the
size of these terms that really matters here, but rather the stability of the
fit to the on-peak, equal mass, $\rho\rho$ contribution, for which the isospin
analysis is to be carried out. If the fits are sensitive to terms of the form
in Eq.~(\ref{fitterm}), then further analysis, and probably significantly more
data is needed.

As an alternative to fitting the data including terms of the form
(\ref{fitterm}), one can eliminate effects of any contributions of this form by
decreasing the width of the $\rho$ bands, $\Delta$, used in the fit (or
imposing a cut on $|m_1-m_2|$). Once the accepted $\rho$ band is small enough,
the result will be stable against further reduction in its width, and also
against changes to the leading fit parameters when a term of the form
(\ref{fitterm}) is added. At present, BaBar uses a band  $0.52\,{\rm GeV} <
m_{\pi\pi} < 1.02\,{\rm GeV}$~\cite{Aubert:2003mm,Aubert:2003zv} whereas BELLE
accepts a narrower range, $0.65\,{\rm GeV} < m_{\pi\pi} <0.89\,{\rm
GeV}$~\cite{Zhang:2003up}. The possible $I=1$ contamination in the
$B\to\rho\rho$ signal diminishes for $\Delta < \Gamma_\rho$ at least as
$(\Delta/m_\rho)^2$. If the extracted values of the rates are stable for
different values of $\Delta$ that would indicate that the $I=1$ contamination
is small and we need not worry further about these types of terms, whereas
results that are sensitive to $\Delta$ would indicate that there is a
contribution of this type that must be more carefully investigated, or excluded
by taking a smaller acceptance.

Clearly both the approach of adding parameters to the fit and the approach of
narrowing the acceptance have a statistical cost. We are hopeful that, even
with the present data set, one will be able to see that the impact of possible
$I=1$ terms is not large. If their effect turns out to be important, then more
data will be needed to eliminate their impact. 

%%%%%%%%%%%%%%%%%%%%%%%%%%%%%%
\section{Isospin Relations}

For each transversity, $\sigma$, the even-isospin amplitudes have relationships
similar to that for the two-pion amplitudes~\cite{Gronau:1990ka},
\beqa\label{triangle}
\frac{1}{\sqrt2}\,A^{\sigma}_{+-}+A^{\sigma}_{00}
  &=&A^{\sigma}_{+0}\,,\no\\
\frac{1}{\sqrt2}\,\bar A^{\sigma}_{+-}+\bar A^{\sigma}_{00}
  &=&\bar A^{\sigma}_{-0}\,.
\eeqa
Each of these equations can be represented as a triangle in the complex plane.
Note that the triangles corresponding to the different transversity states can
be different.

Tree diagrams contribute to both $\Delta I=1/2$ and $3/2$ transitions to $I=0$
and $I=2$ final states, respectively. Since the gluon is isospin singlet,
penguin diagrams contribute only to $\Delta I=1/2$ transitions to $I=0$ final
states. Since the final $\rho^\pm\rho^0$ states have no $I=0$ component,
$A^\sigma_{+0}$ and $\bar A^\sigma_{-0}$ are pure tree amplitudes.  Therefore,
$|A^{\sigma}_{+0}|=|\bar A^{\sigma}_{-0}|$ and the relative phase of these
amplitudes is $2\gamma$ [see Eq.~(\ref{defpt})]. The two triangles originating
from Eqs.~(\ref{triangle}) for any given $\sigma$ can thus be superimposed with
a common base, $A^\sigma_{+0}$, if all the $\bar A_{ij}^\sigma$ amplitudes are
multiplied by a factor $e^{2i\gamma}$.

Electroweak penguin amplitudes, unlike gluonic penguins, contribute to both
$\Delta I=1/2$ and $3/2$ and hence cannot be  distinguished from the tree
amplitudes by their isospin structure. Since electroweak penguins contribute to
both $T_{ij}$ and $P_{ij}$ in Eq.~(\ref{defpt}), one impact of such terms would
be a possible difference between $A_{+0}$ and $e^{2i\gamma}\bar A_{-0}$.  The
size of corrections that contribute to $|A_{+0}| \neq |\bar A_{-0}|$ can be
constrained by measuring these two rates. The average of the
BABAR~\cite{Aubert:2003mm} and BELLE~\cite{Zhang:2003up} results is
\beqa\label{chargedrates}
{\cal A}_{\mp0}&=&\frac{|\bar A_{-0}|^2-|A_{+0}|^2}{|\bar
  A_{-0}|^2+|A_{+0}|^2}=-0.09\pm0.16\,,\no\\
f_0&=&\frac{|\bar A^0_{-0}|^2+|A^0_{+0}|^2}{|\bar
  A_{-0}|^2+|A_{+0}|^2}=0.96^{+0.04}_{-0.06}\,.
\eeqa
These results are consistent with the isospin relationship ${\cal A}_{\mp0}=0$,
though with current precision the test is not particularly stringent. Given
this, there is residual uncertainty in the extracted value of $\alpha$ that is
not constrained by the isospin analysis.  While the impact of electroweak
penguins on the extraction of $\alpha$ from the isospin analysis of
$B\to\pi\pi$ can be estimated to be of order $1.5^\circ$~\cite{Gronau:1998fn},
estimates of other isospin violating effects employ hadronic models and range
from negligible to less than $5^\circ - 10^\circ$~\cite{ciuchini}.  The impact
of these effects is expected to be similar in $B\to \rho\rho$.  Dedicated
analyses are warranted, since both the matrix elements of electroweak penguin
operators and isospin breaking are different for the $\rho\rho$ final state. 
At the present level of accuracy it is reasonable to assume that these
uncertainties are small compared to those that are bounded by the isospin
analysis; we will neglect them in what follows.

Once the branching ratios ${\cal B}[B\to(\rho^i\rho^j)_\sigma] =
|A_{ij}^\sigma|^2$ are measured, one can construct the two triangles and  use
this construction to measure the relative phase between $A^{\sigma}_{+-}$ and
$e^{2i\gamma}\bar A^{\sigma}_{+-}$~\cite{Gronau:1990ka}. This phase is
$2\delta_\sigma$ defined in Eq.~(\ref{comrel}).  It arises from a combination
of relative weak and strong phases and the relative magnitudes of the $T_{+-}$
and $P_{+-}$ contributions, none of which can be reliably calculated.  Using
the two-triangle construction to determine $2\delta_\sigma$, there is a
fourfold ambiguity in the value of this phase, coming from the four possible
orientations of the two triangles relative to their common base.

Until the flavor-tagged branching fractions, ${\cal B}
[B^0\to(\rho^0\rho^0)_\sigma]$ and ${\cal B}
[\overline{B}{}^0\to(\rho^0\rho^0)_\sigma]$, are separately measured, one
cannot determine $\delta_\sigma$. However, one can bound it.  Among the three
averaged branching ratios (summed over transversities),
\beqa\label{avebr}
{\cal B}_{+-}&=&\frac12\left(|A_{+-}|^2+|\bar A_{+-}|^2\right),\no\\
{\cal B}_{+0}&=&\frac12\left(|A_{+0}|^2+|\bar A_{-0}|^2\right),\no\\
{\cal B}_{00}&=&\frac12\left(|A_{00}|^2+|\bar A_{00}|^2\right),
\eeqa
the first two have been measured and there is an upper bound on the third. This
provides an upper bound on ${\cal B}^\sigma_{00}$ for any $\sigma$. It is
significantly smaller than the rate for the dominant longitudinal mode in the
other channels.  This allows us to place a significant bound on $\delta_0$,
using the construction described above. Explicitly, the bound
reads~\cite{Grossman:1997jr,Gronau:2001ff}\footnote{The bound
in~\cite{Gronau:2001ff}, quoted in Eq.~(\ref{bounds}), is the same as the one
in Eq.~(2.15) of~\cite{Grossman:1997jr} up to terms of ${\cal O}[({\cal
B}_{00}/{\cal B}_{+0})^2,({\cal B}_{+-}/{\cal B}_{+0}-2)^2]$,
where~\cite{Gronau:2001ff} is more restrictive. In~\cite{Gronau:2001ff}, the
weaker bound in Eq.~(2.12) of~\cite{Grossman:1997jr},
$\cos2\delta_0\geq1-2{\cal B}_{00}^0/{\cal B}_{+0}^0$, is referred to as the
Grossman-Quinn bound.}
\beq\label{bounds}
\cos2\delta_0\geq1-\frac{2{\cal B}_{00}^0}{{\cal B}_{+0}^0}
  +\frac{({\cal B}_{+-}^0-2{\cal B}_{+0}^0+2{\cal B}_{00})^2}{4{\cal
    B}_{+-}^0{\cal B}_{+0}^0}\,.
\eeq
This bound can be further strengthened if experiments constrain
$C_{+-}^0$~\cite{Charles:1998qx} and $C_{00}^0$.

%%%%%%%%%%%%%%%%%%%%%%%%%%%%%%%%
\section{\boldmath Corrections proportional to $1-f_0$}

For both $B\to\rho^+\rho^-$ and $B\to\rho^\pm\rho^0$, experiments have
determined that the longitudinal fraction $f_0$ is close to $100\%$ [see
Eq.~(\ref{twocharged})]. Thus, even if the experiments do not distinguish the
asymmetry in the longitudinal mode alone, one can use the total asymmetry to
constrain the longitudinal asymmetry. Since we already know from the data that
the decay is almost purely longitudinal, the correction is small, of ${\cal
O}(1-f_0)$. Using $S_{+-}=\sum_\sigma f_\sigma S_{+-}^\sigma$ and
$C_{+-}=\sum_\sigma f_\sigma C_{+-}^\sigma$, the differences between the
transversity-summed CP violating asymmetries and those in the longitudinal mode
are given by
\beqa 
S_{+-}^0-S_{+-}&=&(1-f_0)\Bigg(S_{+-}^0
  -\frac{S_{+-}^\parallel+S_{+-}^\perp}{2}\Bigg)
  -\left(f_\parallel-f_\perp\right)
  \frac{S_{+-}^\parallel-S_{+-}^\perp}{2}\,, \no\\
C_{+-}^0-C_{+-}&=&(1-f_0)\Bigg(C_{+-}^0
  -\frac{C_{+-}^\parallel+C_{+-}^\perp}{2}\Bigg)
  -\left(f_\parallel-f_\perp\right)
  \frac{C_{+-}^\parallel-C_{+-}^\perp}{2}\,.
\eeqa
The $S_{+-}^\sigma$ and $C_{+-}^\sigma$ asymmetries in each of the transversity
channels can in principle be anywhere from $-1$ to $+1$ subject to the
constraints $(S_{+-}^\sigma)^2 + (C_{+-}^\sigma)^2 \leq 1$. Thus, the maximal
deviations of the measured asymmetries from those for the longitudinal modes
are
\beqa\label{maxiss}
|S_{+-}^0-S_{+-}|&\leq&(1-f_0)\left(1+|S_{+-}^0|\right), \no\\
|C_{+-}^0-C_{+-}|&\leq&(1-f_0)\left(1+|C_{+-}^0|\right).
\eeqa
In reality, we expect the error in estimating $S_{+-}^0$ to be smaller than
this upper bound. To zeroth order in $|P_{+-}^\sigma/T_{+-}^\sigma|$ we have
$S_{+-}^\parallel=-S_{+-}^\perp=S_{+-}^0$. Consequently, we obtain
\beq\label{realss}
S_{+-}^0-S_{+-} = (1-f_0-f_\parallel+f_\perp)\, S_{+-}^0
  + {\cal O}[ (1-f_0)\, |P_{+-}/T_{+-}|\;\! ] \,.
\eeq

One further issue that must be considered is the impact of non-resonant
contributions to $B$ meson decays to four pions, and that of other resonances
that yield the same final state in this analysis. These could contribute with
opposite CP to that of the dominant longitudinal mode. Since the angular
distribution given by the decay of a spin-1 longitudinally-polarized meson is
quite restrictive, the contamination due to all such contributions is
effectively included in the error of $1-f_0$, the fraction of non-longitudinal
contributions.  Thus the uncertainty due to these contributions is taken into
account by allowing for the uncertainties in Eq.~(\ref{maxiss}) when
determining the CP asymmetries in the longitudinal mode.

%%%%%%%%%%%%%%%%%%%%%%%%%%%%%%%%%%%%
\section{Numerical results}

The experimental values given by BABAR for the three averaged branching ratios
defined in Eq.~(\ref{avebr}) are~\cite{Aubert:2003mm,Aubert:2003zv}
\beqa \label{twocharged}
{\cal B}_{+-}&=&(27^{+7+5}_{-6-7})\times10^{-6}\,,\qquad\qquad\ \
(f_0)_{+-}=0.99^{+0.01}_{-0.07}\pm0.03\,,\no\\
{\cal B}_{+0}&=&(22.5^{+5.7}_{-5.4}\pm5.8)\times10^{-6}\,,\qquad
(f_0)_{+0}=0.97^{+0.03}_{-0.07}\pm0.04\,,\no\\
{\cal B}_{00}&<&2.1\times10^{-6} \quad (90\% \mbox{ CL})\,,
\eeqa
while BELLE obtained~\cite{Zhang:2003up}
\beq\label{belres}
{\cal B}_{+0}=(31.7\pm7.1^{+3.8}_{-6.7})\times10^{-6},\qquad
(f_0)_{+0}=0.948\pm0.106\pm0.021\,.
\eeq

We take ${\cal B}_{+-} = {\cal B}_{+-}^0$ and ${\cal B}_{+0} = {\cal B}_{+0}^0$,
thus introducing errors of order $(1-f_0)$. These are much smaller than the
present experimental errors on ${\cal B}_{+-}$ and ${\cal B}_{+0}$ and
therefore can be neglected. We use the following averages, based on
(\ref{twocharged}) and (\ref{belres}):
\beqa \label{avbrex}
{\cal B}_{+-}^0&=&(27\pm9)\times10^{-6},\no\\
{\cal B}_{+0}^0&=&(26\pm6)\times10^{-6},\no\\
{\cal B}_{00}^0&=&(0.6^{+0.8}_{-0.6})\times10^{-6}.
\eeqa
The value of ${\cal B}_{00}$ is based on scaling the number of signal events
given in Ref.~\cite{Aubert:2003mm} and conservatively assuming the efficiency
for $(f_0)_{00} = 1$, which yields the largest rate~\cite{Andrei}.

The first question to be asked is whether the rates in Eq.~(\ref{avbrex}) are
consistent with isospin symmetry. Note that in the ${\cal B}_{00}^0\to 0$
limit, we must have ${\cal B}_{+-}^0=2{\cal B}_{+0}^0$. The central values in
(\ref{avbrex}) imply a small ${\cal B}_{00}^0$ but ${\cal B}_{+-}^0\sim {\cal
B}_{+0}^0$, thus the consistency with the isospin constraints is limited.
Indeed, a statistical analysis~\cite{Hocker:2001xe} of the rates in
Eq.~(\ref{avbrex}) finds the goodness of the fit is only 24\%.  Since this
confidence level is not extremely small, in the following we derive limits on
$\delta_0$ assuming that isospin symmetry holds.\footnote{It was pointed out in
Ref.~\cite{Aubert:2003zv} that the small upper bound on ${\cal B}_{00}/{\cal
B}_{+0}$ constrains the penguin pollution.}  Using the isospin constraints as
coded in~\cite{Hocker:2001xe} and the branching ratios in Eq.~(\ref{avbrex}),
we obtain the 90\% CL bound:
\beq\label{grqunu}
\cos2\delta_0 > 0.83\,,
\eeq
or, equivalently,
\beq\label{twodelnu}
|\delta_0| < 17^\circ.
\eeq
Note that even though the statistical significance of ${\cal B}_{+-}^0 
- 2{\cal B}_{+0}^0 + 2{\cal B}_{00} \neq 0$ is small, the last term in
Eq.~(\ref{bounds}) does play a role.  Had we ignored it, we would have 
obtained $\cos2\delta_0 > 0.80$.

It is interesting that the small value of ${\cal B}_{00}/{\cal B}_{+0}$ already
puts an upper bound on $C_{+-}$, the measure of direct CP violation. For each
transversity component, the isospin relations imply, for ${\cal B}_{00}^\sigma
/ {\cal B}_{+0}^\sigma<1/2$,
\beq\label{maxcpm}
|C_{+-}^\sigma| < 2\, \sqrt{
  \frac{{\cal B}_{00}^\sigma}{{\cal B}_{+0}^\sigma} - \left(
  \frac{{\cal B}_{00}^\sigma}{{\cal B}_{+0}^\sigma} \right)^2}\,.
\eeq
The 90\% CL bound on ${\cal B}_{00}^0/{\cal B}_{+-}^0$ that can be
extracted from Eq.~(\ref{avbrex}) yields, to leading order in the
small quantity $1-f_0$,
\beq\label{maxcnu}
|C_{+-}^0| < 0.53\,.
\eeq

%%%%%%%%%%%%%%%%%%%%%
\section{Conclusions}

The present measurements of the rates of the various $B\to\rho \rho$ decays
already yield significant limits on the uncertainty in the extraction of
$\alpha$ from the CP violating asymmetry in $B^0$ and $\overline{B}{}^0$ decays
to $\rho^+\rho^-$.  Given the large branching fractions of these channels, we
look forward to an asymmetry measurement in the near future which will
determine $\alpha$ with interesting precision. To ensure the accuracy of the
results it is important to include an isospin-1 contribution in the fits to
data, as in Eq.~(\ref{fitterm}), constrained to vanish when the two $\rho$
mesons have equal masses. We do not expect the impact of this contribution to
be large, but it could introduce changes of order $(\Gamma_\rho/m_\rho)^2$ to
the best fit parameters. Once this effect is constrained experimentally and the
CP-violating quantity $S_{+-}$ is measured, $B\to\rho\rho$ decays promise to
provide the best model independent determination of the parameter $\alpha$ for
some time to come.

%%%%%%%%%%%%
\acknowledgments
We thank Y.~Frishman, J.~Fry, W.~Goldberger, A.~Gritsan, Y.~Grossman,
S.~Laplace, G.~Raz, A.~Schwimmer and T.~Volansky for useful discussions.
We thank the Aspen Center for Physics for hospitality, that is where this work
was initiated.
The work of A.F. was supported in part by the U.S. National Science
Foundation under grant NSF-PHY-9970781.
Z.L.\ was supported in part by the Director, Office of Science, Office of
High Energy and Nuclear Physics, Division of High Energy Physics, of the  U.S.\
Department of Energy under Contract DE-AC03-76SF00098 and by a DOE Outstanding
Junior Investigator award.
Y.N.\ is supported by the Israel Science Foundation founded by the Israel
Academy of Sciences and Humanities, by EEC RTN contract HPRN-CT-00292-2002,
by a Grant from the G.I.F., the German-Israeli Foundation for Scientific
Research and Development, and by a grant from the United States-Israel
Binational Science Foundation (BSF), Jerusalem, Israel.
The work of H.Q.\ was supported by the Department of Energy, contract
DE-AC03-76SF00515.

%%%%%%%%%%%%%%%%%%%%%

\end{document}